# Model-based determination of the synchronization delay between MRI and trajectory data

P. I. Dubovan[1,2] and C. A. Baron* [1,2]

[1] Department of Medical Biophysics, Western University, London Ontario Canada
[2] Centre for Functional and Metabolic Mapping, Western University, London Ontario Canada

* Corresponding author: Corey A. Baron – corey.baron@uwo.ca

Abstract

**Purpose**: Real time monitoring of dynamic magnetic fields has recently become a commercially available option for measuring MRI k-space trajectories and magnetic fields induced by eddy currents in real time. However, for accurate image reconstructions, sub-microsecond synchronization between the MRI and trajectory data is required. In this work, we introduce a new model-based algorithm to automatically perform this synchronization using only the MRI and trajectory data.

**Methods**: The algorithm works by enforcing consistency between the MRI data, trajectory data, and receiver sensitivity profiles by iteratively alternating between convex optimizations for (a) the image and (b) the synchronization delay. A healthy human subject was scanned at 7T using a transmit-receive coil with integrated field probes using both single shot spiral and echo-planar imaging (EPI), and reconstructions with various synchronization delays were compared to the result of the proposed algorithm. The accuracy of the algorithm was also investigated using simulations, where the acquisition delays for simulated acquisitions were determined using the proposed algorithm and compared to the known ground truth.

**Results**: In the *in vivo* scans, the proposed algorithm minimized artefacts related to synchronization delay for both spiral and EPI acquisitions, and the computation time required was less than 30 seconds. The simulations demonstrated accuracy to within tens of nanoseconds.

**Conclusion**: The proposed algorithm can automatically determine synchronization delays between MRI data and trajectories measured using a field probe system.

## 1 Introduction

Non-cartesian MRI provides the advantages of increased SNR due to shortened echo times, robustness to motion, and more time-efficient data acquisition. However, non-Cartesian approaches are generally more sensitive to k-space errors from eddy currents compared to Cartesian sampling, which ultimately results in blurring and/or geometric distortions. Many approaches to estimate these k-space errors have been proposed, which include direct measurement of the trajectory (1–3), calibration based on a gradient impulse response function (4,5), or via retrospective modelling (6). Alternatively, real time monitoring of eddy current fields using field probes has recently become a commercially available rapid option for measuring k-space trajectories and eddy current fields in real time, up to 2nd or 3rd order in space (7–10). These measured field dynamics can be included together with an off-resonance map in an expanded encoding model based reconstruction that greatly ameliorates artefacts from off-resonance and eddy currents (7). However due to hardware specific filter delays that differ between the MRI and field-probe spectrometers, the relative timing between the field-probe measured trajectory and the MRI signal is unknown. Errors in this timing will henceforth be referred to as a "synchronization delay".

Gradient delays and their associated artefacts have been well-described for non-Cartesian acquisitions that do not utilize external field monitoring, where there are generally separate delays for each of the three gradient channels instead of the single global synchronization delay that is required to be determined for field-monitored acquisitions. Various approaches for correcting for these delays have been proposed, which include pulse sequence modifications or a separate calibration scan (11–13) and retrospective methods that are designed for particular trajectories (11,14–17). While these methods could likely be adapted to determine the synchronization delay required for expanded encoding model acquisitions, the necessity for either specific pulse sequence prescans or specific trajectories creates a complicated solution landscape with various trade-offs depending on the approach. A promising self-consistency approach that explicitly determines delays has been proposed, where gradient delays and receiver sensitivity profiles are simultaneously estimated from fully sampled calibration data using low-rank constraints (18), but this method is not applicable for situations where the receiver sensitivity calibration data is obtained from a separate scan, as is typically the case for reconstructions using field-probe measurements.

In this work, we introduce a general model-based retrospective approach to determine the synchronization delay between a measured trajectory and the MRI data that is applicable to arbitrary trajectories and requires no pulse sequence modification. Further, we demonstrate that it can accurately estimate the delay even when aggressive coil compression is utilized and propose an approach to greatly accelerate convergence, making this method require very little time for computations. We demonstrate the performance of the algorithm for both single-shot spiral and EPI acquisitions in the brain of a healthy human subject and through simulation.

## 1.1 Theory

The m'th signal $y_m^j$ measured at the n'th location $r_n$ and time $t_m$ by receiver element $j$ can be modelled using (7,19):

$$y_m^j = \sum_n C^j(r_n)x(r_n)e^{i\omega(r_n)t_m}e^{i\sum_l \ k_l(t_m+\tau+\Delta\tau)b_l(r_n)}, \quad (1)$$

where $C^j$ is the j'th receiver sensitivity profile, x is the image, $k_l$ are the coefficients for the spherical harmonic basis functions $b_l$ ($l$ indexes the different spherical harmonic terms) measured by a field probe system, $\omega$ is the off-resonance from $B_0$ inhomogeneity (measured in a separate scan), $\tau$ is the presumed synchronization delay of the MRI signal $y$ relative to $k_l(t)$, and $\Delta\tau$ is the unknown error in $\tau$. $\tau$ might be initially estimated based on typical trigger delays or the time for the gradient prephasers required for some trajectories (e.g., EPI), for example, and the purpose of this algorithm is to determine the net error-free delay, $\tau + \Delta\tau$. To cast Equation 1 into a form that enables a convex estimation of $\Delta\tau$, we first recognize that for typical delay errors that are on the order of microseconds, $k_l(t_m)$ is approximately linear in the vicinity of each sample $m$. This approximation allows the substitution $k_l(t_m + \tau + \Delta\tau) = k_l(t_m + \tau) + \Delta\tau k_l'(t_m + \tau)$ where $k_l'(t) = dk_l(t)/dt$, yielding

$$y_m^j = \sum_n C^j(r_n)x(r_n)e^{i\omega(r_n)t_m}e^{i\sum_l \ k_l(t_m+\tau)b_l(r_n)}e^{i\Delta\tau\sum_l \ k_l'(t_m+\tau)b_l(r_n)}. \quad (2)$$

Further, for small $\Delta\tau$, the net phase $\Delta\tau\sum_l \ k_l'(t_m + \tau)b_l(r_n) << 1$, allowing the use of a Taylor approximation $e^{i\Delta\tau\sum_l k_l'(t_m+\tau)b_l(r_n)} \approx 1 + i\Delta\tau\sum_l \ k_l'(t_m + \tau)b_l(r_n)$, yielding:

$$y_m^j = \sum_n C^j(r_n)x(r_n)e^{i\omega(r_n)t_m}e^{i\sum_l \ k_l(t_m+\tau)b_l(r_n)} + \Delta\tau i\sum_n\sum_l k_l'(t_m + \tau)b_l(r_n)C^j(r_n)x(r_n)e^{i\omega(r_n)t_m}e^{i\sum_l \ k_l'(t_m+\tau)b_l(r_n)}. \quad (3)$$

Equation 3 can be cast into a matrix equation over all receivers:

$$\Delta\tau Bx = Y - Ax \tag{4}$$

$$Y_{m+jN_s} = y_m^j$$

$$A_{m+jN_s,n} = C^j(r_n)e^{i\omega(r_n)t_m}e^{i\sum_l k_l(t_m+\tau)b_l(r_n)}$$

$$B_{m+jN_s,n} = i\sum_l k_l'(t_m+\tau)b_l(r_n)C^j(r_n)x(r_n)e^{i\omega(r_n)t_m}e^{i\sum_l k_l'(t_m+\tau)b_l(r_n)}$$

where $N_s$ is the number of samples acquired. If $N_r$ is the number of receivers and $N_x$ is the number of object-domain voxels in $x$, then A and B are matrices of size $N_sN_r$ by $N_x$. Accordingly, $Bx$ and $Y - Ax$ are each column vectors with length equal to $N_sN_r$, and $\Delta\tau$ can be determined in a least squares sense via

$$\Delta\tau = Re[((Bx)^H(Bx))^{-1}(Bx)^H(Y - Ax)]. \tag{5}$$

The determination of $\Delta\tau$ requires knowledge of the image x. Accordingly, we jointly solve for x and $\Delta\tau$ by iteratively alternating between determining $x$ at iteration $p$ using

$$x_p = argmin_x \left\|A_p x - Y\right\|_2^2, \tag{6}$$

and solving for $\Delta\tau$ using direct matrix implementation of Equation 5. Equation 6 is solved using the conjugate gradient method. For every iteration, $p$, $k_l(t_m + \tau_p)$ is re-interpolated from the original field probe samples via $\tau_p = \tau_{p-1} + \Delta\tau_{p-1}$. Likewise, $k_l'(t)$ is estimated using a 5-point central difference of the field probe data which is then interpolated to the MRI samples located at $t_m + \tau_p$. All interpolations used piecewise cubic Hermite interpolation (20). Iterations continue until $\Delta\tau_p$ is less than a user-specified threshold, $\Delta\tau_{min}$. The net algorithm, Algorithm 1, is portrayed in Table 1. Notably, this joint optimization is only required for a single sample slice because the delay is not expected to change for different slices.

In practice, Algorithm 1 may converge slowly. To accelerate convergence, we propose to scale $\Delta\tau_p$ by a factor $\gamma$ in every iteration. If the polarity of $\Delta\tau_p$ changes from one iteration to the next it suggests that $\gamma$ is too large, and thus $\gamma$ is reduced by a factor of 2 for the next iteration, subject to $\gamma \geq 1$.

Coil compression performs a compression of the receiver channels into fewer virtual channels using a linear transformation determined from singular value decomposition, which can drastically reduce memory requirements and accelerate image reconstructions (21). Accordingly, in this work the impact of coil compression on the accuracy of Algorithm 1 will also be evaluated by varying the number of virtual coils, $N_r$.

## 2 Methods

*2.1 Data Acquisition and Reconstruction Parameters*

A healthy patient was scanned on a 7T head-only MRI (Siemens) at Western University's Centre for Functional and Metabolic Mapping (80 mT/m gradient strength and 400 T/m/s max slew rate), which was approved by the Institutional Review Board at Western University and informed consent was obtained prior to scanning. The scans used a 32 channel receive coil with integrated field probes (22). A single-shot spiral acquisition was performed with an in-plane resolution of 1.5 x 1.5 $mm^2$, 3 mm slice thickness (10 slices), TE/TR = 33/2500 ms, FOV = 192 x 192 $mm^2$, an undersampling rate of 4, and dwell time between samples of 2.5 µs. A single shot spin-echo EPI acquisition was acquired in a separate scanning session with an in-plane resolution of 2 $mm^2$, 3 mm slice thickness (10 slices), TE/TR = 59/2500 ms, FOV = 192 x 192 $mm^2$, an undersampling rate of 3, and dwell time between samples of 2.4 µs. Noise correlation between receivers was corrected using prewhitening before any reconstructions (23). $B_0$ field maps were acquired at the same resolution as the single-shot spiral scan using dual echo gradient echo with TE values of 4.08 ms and 5.10 ms. The first echo of the same data was used to estimate $C^j$ via ESPIRiT (24). The spatially varying field dynamics up to 2nd order in space were measured simultaneously with the MRI data using a field monitoring system (Skope) consisting of 16 transmit/receive 19F field probes and a sampling dwell time of 1 µs (22). In all implementations of Algorithm 1, iterations in $p$ were stopped when $\Delta\tau$ < 0.005 µs. The object domain support (i.e., the FOV images were reconstructed to in Equation 6) was determined by thresholding the $B_0$ mapping images, which optimizes reconstruction time (25).

*2.2 Convergence and Coil Compression Performance*

To investigate the speed increases for $\gamma_0$ > 1, the number of iterations required to determine $\tau$ was measured for two slices of the in vivo spiral data for three cases corresponding to $\gamma_0 = \{1,3,10\}$. Each of these cases did not use coil compression. To investigate the performance of the algorithm with coil compression, $\tau$ was estimated from the same two *in vivo* slices with the full complement of 32 channels as well as 16, 8 and 5 virtual channels ($\gamma_0$ = 3 for all cases). Also, to shed light on the sensitivity of x on the number of virtual channels, the value of $\tau$ determined from all 32 channels was used in Equation 6 to solve for $x$ with fewer virtual channels.

### *2.3 Validation of Accuracy*

Simulations to validate accuracy were performed by sampling the image acquired in the $B_0$ mapping scan using the forward model defined by Equation 1 using measured field probe dynamics with various choices of $\tau$ ranging from -2.5 µs to 10 µs for both the spiral and EPI acquisitions described above. Gaussian white noise (SNR ~ 10 in white matter) was added to the simulated data before performing any reconstructions.

## 3 Results

The field dynamics measured using the field probe system for both spiral and EPI are displayed in Fig. 1, where it is observed that the assumption of slowly varying dynamics is valid on time scales of tens to hundreds of µs.

For the *in vivo* data, the images reconstructed from the proposed algorithm shows good image quality at the automatically determined value of $\tau$ (Fig. 2a). Substantial artefacts are introduced for EPI when errors in delay are present, but not for spiral (Fig. 2b,c).

When $\gamma = 1$, a monotonic convergence of $\tau$ is observed, which is greatly accelerated with $\gamma = 3$ (Fig. 3a). When the initial $\gamma$ is chosen to be much too large at a value of 10, the convergence is oscillatory, but still much faster than when $\gamma = 1$. When coil compression is utilized to determine $\tau$, there is negligible degradation in accuracy of $\tau$ (Fig. 3b).

Substantial noise was added to the simulated image reconstructions for both spiral and EPI trajectories, which provides a scenario that is more challenging for accurate determination of $\tau$ than typical in vivo scenarios (Supporting Information Fig. S1). Nevertheless, highly accurate determination of $\tau$ is observed for all levels of coil compression for both spiral and EPI (Fig. 4).

The loss function, $\left\|A_p x - Y\right\|_2^2$ is shown for a range of $\tau$ in Fig. 5. A local minimum is observed for the EPI trajectory for extremely large errors in $\tau$, which causes Algorithm 1 to fail because the starting guess is in the proximity of the local minimum.

## 4 Discussion

In this work, we have developed a new algorithm for the accurate and rapid determination of the delay between MRI data samples and field dynamics measured by external field probes. Consistent delays were determined for different slices (standard deviation of synchronization delays determined from Algorithm 1 was ~ 40 ns across all slices for both spiral and EPI), which suggests that the delay only needs to be calculated once at the beginning of the reconstruction pipeline for a single sample slice. This single joint fit for $\tau$ and x can be performed with aggressive coil compression because this was shown to result in negligible error in $\tau$. However, the image obtained from this joint fit of $x$ and $\tau$ in the sample slice could have artefacts from coil compression (see Supporting Information Fig. S2), and should thus be discarded before reconstructing all slices via Equation 6 with the automatically determined $\tau$ and a more conservative choice of virtual coils. For our GPU implementation, the total time for determination of $\tau$ via Algorithm 1 was 14 s for spiral and 22 s for EPI when using 10 virtual coils and $\gamma_0 = 3$.

The proposed approach was proven to be effective for both single-shot spiral and EPI trajectories. When only six virtual channels are used, the accuracy in $\tau$ for the spiral case is slightly lower than for the EPI case (Fig. 4), which likely occurs because an error in the delay only causes very subtle artefacts for spiral trajectories that may be difficult for the algorithm to detect in the presence of noise (e.g., Fig. 2). Nevertheless, for six virtual coils where the maximum error in $\tau$ was observed, there are negligible artefacts when using that value of $\tau$ in a reconstruction with the full complement of 32 coils (Supporting Information Fig. S2).

When the initial guess for the delay is far from the true value, several starting guesses for $\tau$ may be required to find the global minimum since Algorithm 1 is not globally convex (e.g., for EPI as in Fig. 5); that said, the sub-problems defined by Equations 5 and 6 are each convex. Notably, over 6 months the optimal synchronization delay has not changed by more than ~1 μs on our system, which suggests multiple starting guesses will usually not be necessary. Also, we have developed a simple algorithm to determine the starting guess of $\tau$ for EPI that has been within 1 μs for all cases tested so far. In short, this method consists of: (a) numerically finding the first zero crossing of the first order field probe profile that corresponds to the readout gradient (e.g., blue profile for EPI in Fig. 1b), and then (b) subtracting half the readout duration from the position of the zero-crossing (the readout duration is available from the raw data header).

In other tests, we have found that this method fails when only a single receiver channel is used (data not shown). Accordingly, the ability to estimate $\tau$ likely stems from consistency between the data and receiver sensitivity profiles (i.e., $C^j$ in (1)). The nature of this approach can be intuitively understood by considering the effect of image delays on the solution of Equation 6. For single-shot spiral, a delay primarily causes a slight rotation of the multichannel image data relative to the coil sensitivities, which increases the residual in the loss function $\|A_p x - Y\|_2^2$ ; accordingly, an optimal choice of $\tau$ from Equation 5 will minimize this inconsistency. Other trajectories would have similar effects. For example, a timing delay for EPI creates ghosting (27) and for a radial acquisition creates streaking artefacts (15), which are both inconsistent with the sensitivity profiles. Moreover, since other less general approaches enforcing self-consistency have been effective for radial trajectories (15), it is likely that Algorithm 1, which is also based on self-consistency, would also be successful for this and likely other trajectories. Notably, the algorithm would not be successful for trajectories where delays only cause a benign phase ramp in the object domain (e.g., basic multishot Cartesian trajectories), because there would be no inconsistency with the receiver sensitivities created by the delay error; however, there is also likely little need to accurately determine the synchronization delay in these cases because the delay would not cause any harmful artefacts.

There are two user-defined parameters for this algorithm: the number of virtual coils and the convergence acceleration parameter $\gamma$. However, even substantial coil compression resulted in negligible errors in $\tau$. The authors recommend a somewhat conservative choice of coils for all cases, such as 8 to 12 virtual coils for a 32 channel receiver. Also, while a suboptimal choice of $\gamma$ lengthens the time for convergence, it has no impact on accuracy aside from the unlikely event that it is so large that it brings the delay to a basin of attraction for a local minimum.

While Algorithm 1 only considers a single synchronization delay, it could easily be adapted to estimate different delays on the different gradient channels (i.e., $\tau$ becomes a vector in Equation 5). This adaptation of the algorithm would be appropriate for acquisitions that do not utilize field probe dynamics. However, for this case it would likely be preferred to account for the relative delays between channels and other trajectory errors from eddy currents via a gradient impulse response function (4). Notably, it is possible that there may still exist a global synchronization delay error even after correction with a gradient impulse response function and, accordingly, Algorithm 1 may be applicable for such situations as well.

## 5 Conclusion

In summary, this work has introduced a new model-based retrospective approach to automatically determine the synchronization delay between a k-space trajectory and MRI data, which obviates the need for pulse sequence modifications or manual tuning to determine delays.

## Acknowledgements

This work was supported by the Natural Sciences and Engineering Research Council of Canada (NSERC) Grant Number RGPIN-2018-05448, Canada First Research Excellence Fund to BrainsCAN, Ontario Graduate Scholarship program and the NSERC PGS-D program.

## Data Availability Statement

The source code utilized for the expanded encoding model reconstructions is publicly available at https://gitlab.com/cfmm/matlab/matmri (26). A demo for Algorithm 1 is included.

## Tables

Table 1: Algorithm for the joint optimization of $\tau$ and x, "Algorithm 1"

**Input**: $\Delta\tau_{min}$, $\tau_0$, $\gamma_0$
**Output**: $\tau$, x
**Initialization**: $\Delta\tau_0 = \infty$, $p = 0$
**while** $\Delta\tau_p > \Delta\tau_{min}$ **do**
    $p = p + 1$
    interpolate $k_l(t)$ and $k_l'(t)$ to time samples $t_m + \tau_{p-1}$, and form $A_p$ and $B_p$
    $x_p = argmin_x \left\| A_p x - Y \right\|_2^2$
    $\Delta\tau_p = Re[((B_p x_p)^H (B_p x_p))^{-1} (B_p x_p)^H (Y - A_p x_p)]$
    **if** $\boldsymbol{p} > \mathbf{1}$ and $sign(\Delta\tau_p) \neq sign(\Delta\tau_{p-1})$ **then**
        $\gamma_p = max(1, \gamma_{p-1}/2)$
    **else**
        $\gamma_p = \gamma_{p-1}$
    **end if**
    $\tau_p = \tau_{p-1} + \gamma_p \Delta\tau$
**end while**
**return** $\tau_p$, $x_p$

## Figure Captions

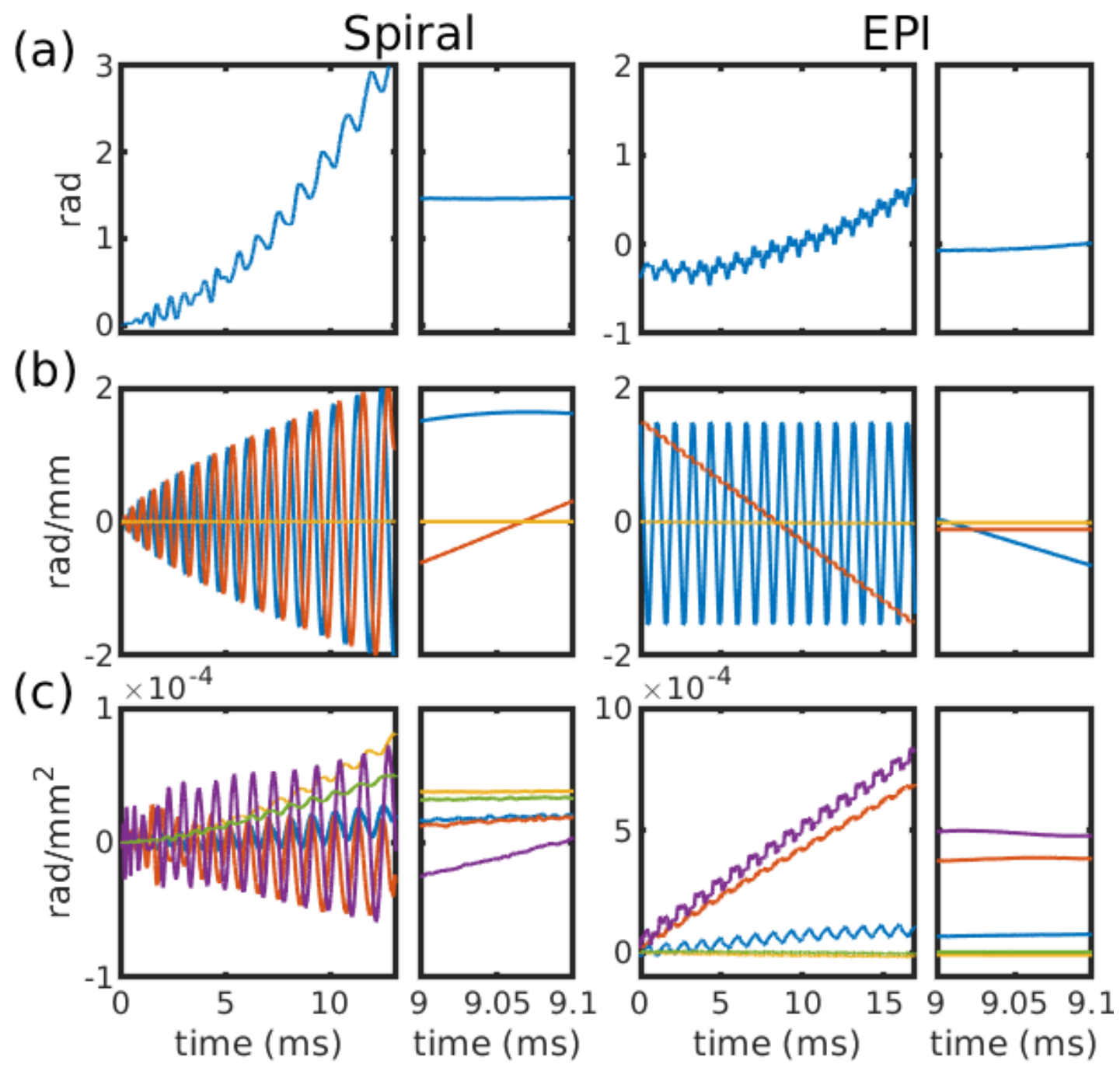

Fig. 1. Field dynamics measured using the field probe system displayed as the coefficients, $k_l(t)$, corresponding to 0th (a), 1st (b), and 2nd (c) order spherical harmonics for each of the spiral and EPI

trajectories. The zoomed-in section highlights a 100 μs span of time, where little variation in fields is observed.

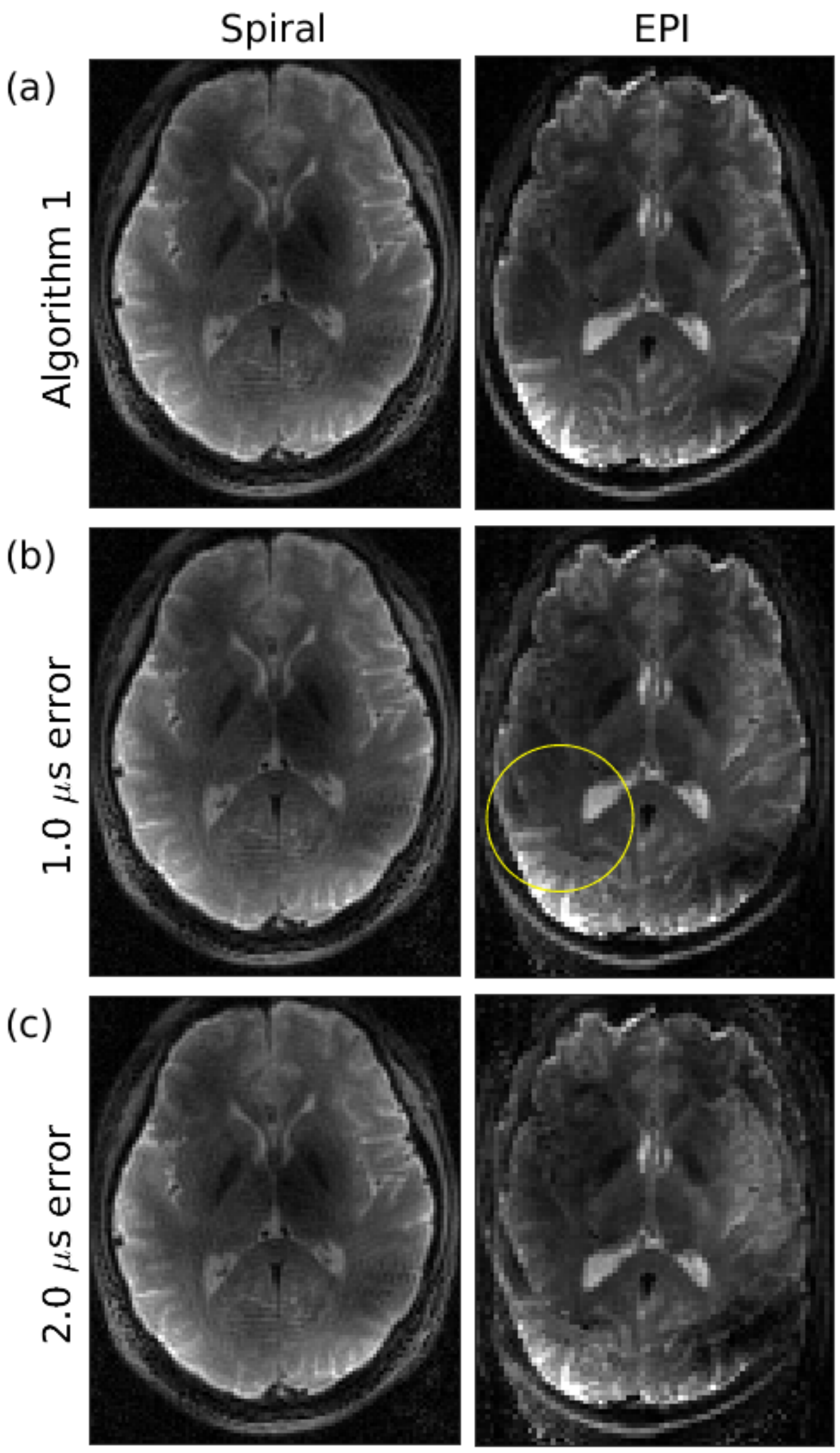

Fig. 2. (a) Image reconstructions that result from Algorithm 1 from both spiral and EPI acquisitions. Panels (b) and (c) show the solution of Equation 6 when $\tau$ is offset from the result in (a) by 1 μs and 2 μs, respectively. While spiral is relatively robust against delay errors, EPI shows artefacts for 1 μs of synchronization error (e.g., circled) and complete reconstruction failure for an error of 2 μs. Similar results are observed in other slices.

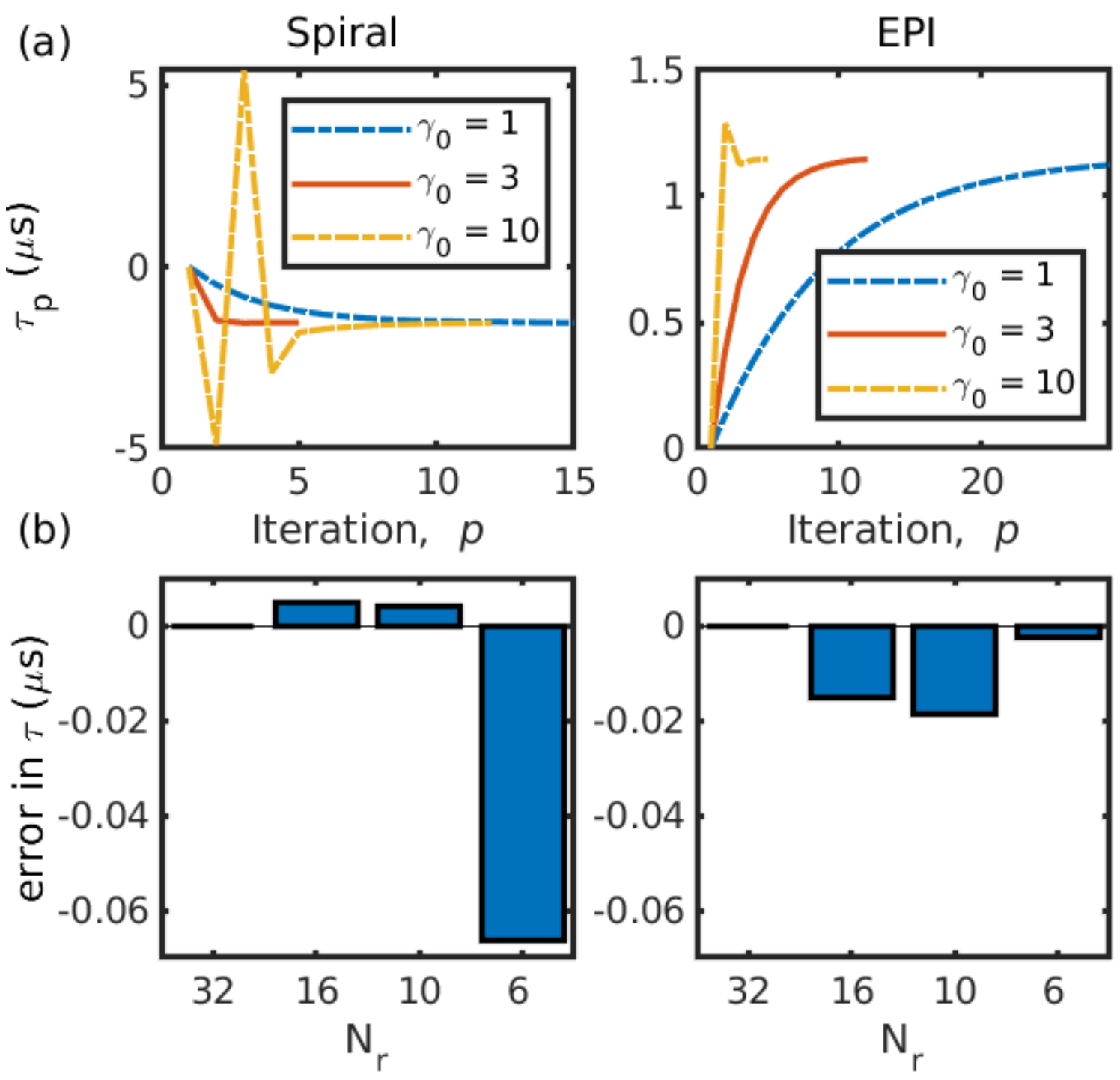

Fig. 3. (a) Convergence of $\tau$ for various choices of $\gamma_0$, where $p$ is the iteration number, for both spiral and EPI prospective scans (results from the same slices shown in Fig. 2a). (b) Error in $\tau$ determined from Algorithm 1 as the number of virtual coils used in coil compression is decreased from the full number of 32 coils, where the result from 32 coils is used as the reference. For both spiral and EPI, negligible error is observed for as low as 6 virtual coils (< 100 ns). Similar results are observed in other slices.

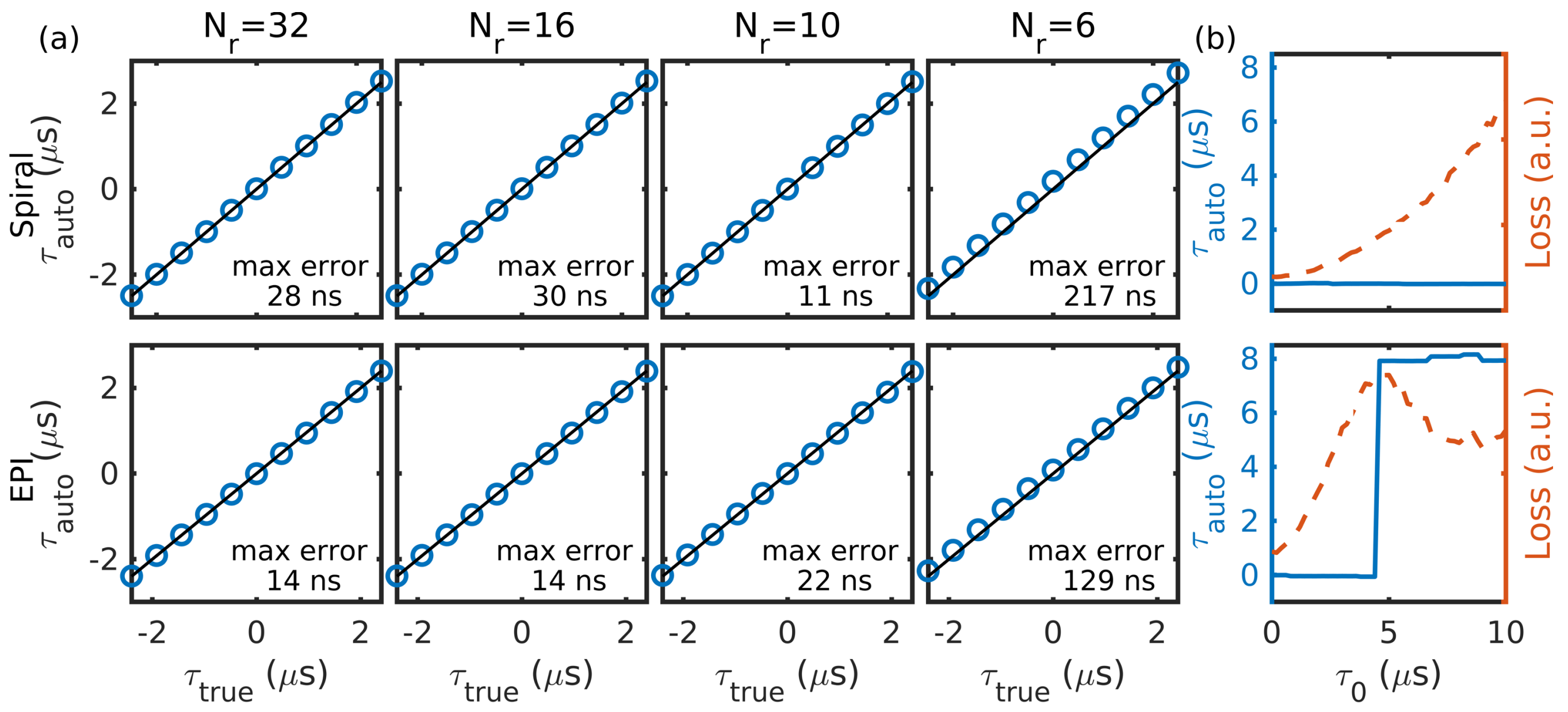

Fig. 4. Simulations when various synchronization delays ($\tau_{true}$) are used to simulate data sampling via Equation 1. After sampling, noise is added to the simulated signal data (SNR ~ 10) and Algorithm 1 is used to estimate the delay, $\tau_{auto}$. For all cases, $\tau_{auto}$ was found using an initial guess $\tau_0 = 0$ µs. Example images from the simulations are shown in Supporting Information Figure S1. (a) For all tested numbers of virtual coils ($N_r$), Algorithm 1 results in accurate estimations of the delay, which mirror the *in vivo* results shown in Figure 2 and 3. (b) Loss function (dashed, right vertical axis) and the error in the result of Algorithm 1 (solid, left vertical axis), where $\tau_0$ is the initial guess used in Algorithm 1. The loss function was computed as $||Ax - Y||_2^2$, where A used $\tau = \tau_0$ (see Equation 4) and $x$ was found via Equation 6 assuming $\tau = \tau_0$, and the true value of $\tau$ was 0 µs. For an initial guess of the synchronization delay more than 5 µs from the true value, Algorithm 1 is attracted to a local minimum for EPI.

**Supporting Information**

Additional supporting information may be found online in the Supporting Information Section.

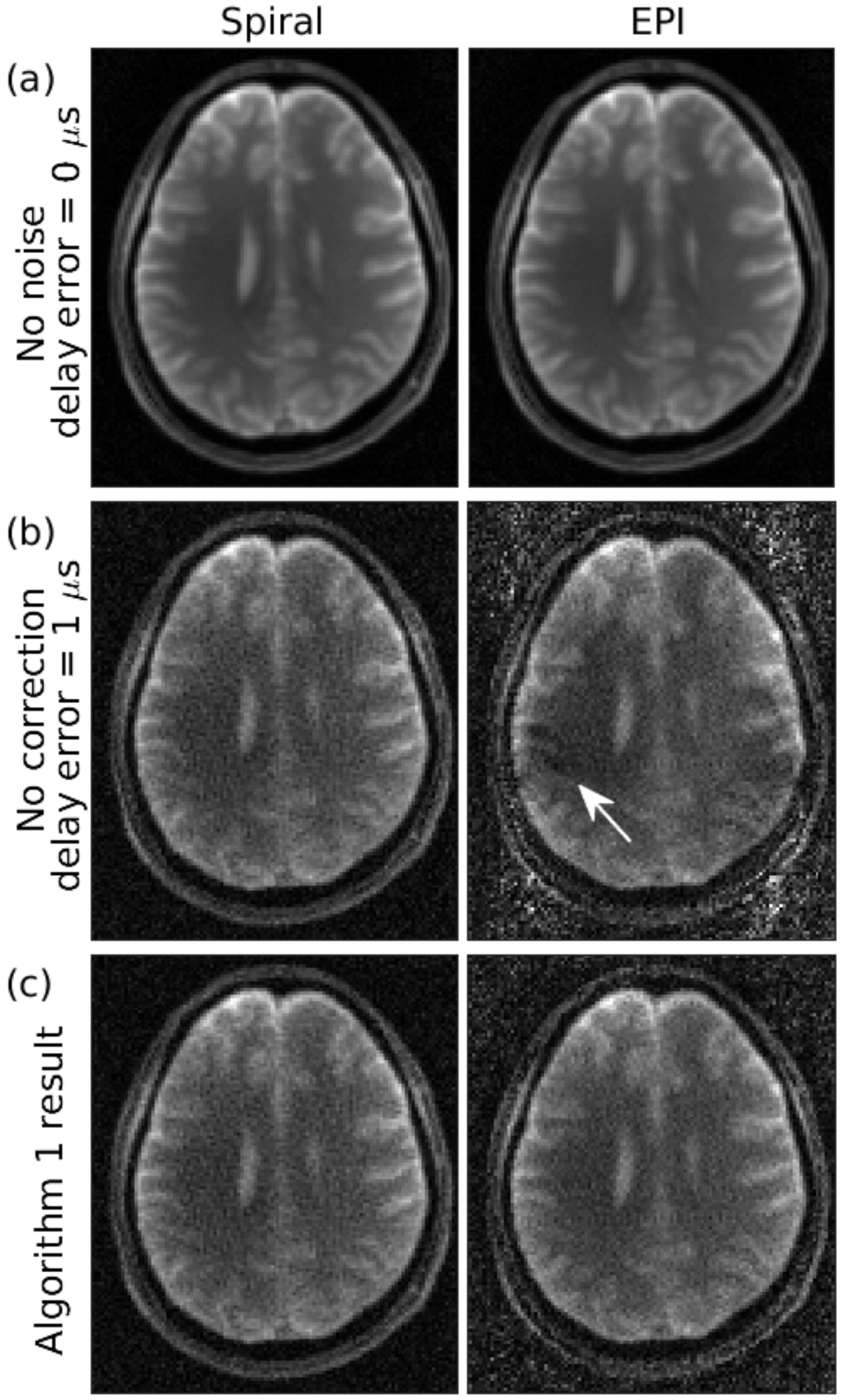

Supporting Information Figure S1. Simulation results for no noise (a), noise added (SNR ~ 10) for 1 μs error in $\tau$ (b), and the result of using Algorithm 1 for noise added (SNR ~ 10) and an initial error in $\tau$ of 1 μs (c). Simulated data, Y, was determined from Equation 1 using an image from the $B_0$ mapping scan as $x$. No artefacts are visible for spiral, while noticeable artefacts are observed for EPI (e.g., white arrow). $N_r$ = 32 for all cases.

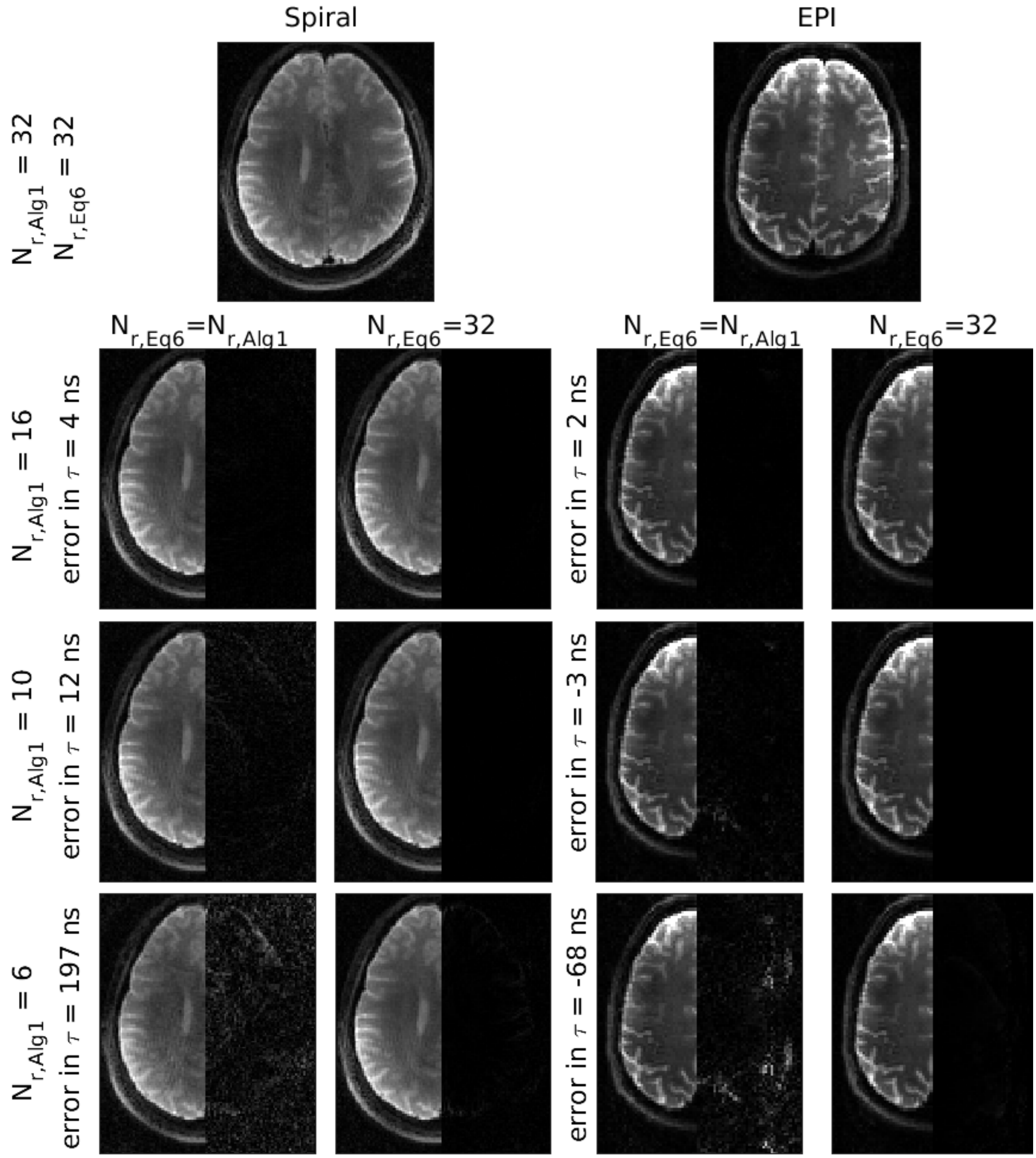

Supporting Information Figure S2. Image reconstructions performed using *in vivo* data in a two step process, where in step 1 the synchronization delay $\tau$ is determined via Algorithm 1 using $N_{r,Alg1}$ virtual coils, and then in step 2 that value of $\tau$ is used in Equation 6 with $N_{r,Eq6}$ virtual coils to determine the final image. The right half of each image shows the difference from the reference image in the top tow, scaled by a factor of 5. The error in $\tau$ is computed as the difference in $\tau$ relative to the case where all 32 coils were utilized.

While the solution to Equation 6 exhibits errors for a low number of virtual coils, Algorithm 1 produces accurate estimations of $\tau$ for all cases. Similar results are obtained in other slices.